\documentclass[aps,prl,twocolumn,superscriptaddress,twoside,floatfix,amsmath,showpacs,nofootinbib]{revtex4}

\usepackage{graphicx}
\usepackage{amsmath}
\usepackage{textcomp}
\usepackage{dcolumn}
\usepackage{bm}

\usepackage{amssymb}
\usepackage[latin1]{inputenc}
\usepackage{amsfonts}
\usepackage{lineno}
\usepackage{url}
\usepackage{array}
\usepackage{verbatim}
\usepackage{subfigure}
\usepackage{listings}
\usepackage{color}
\usepackage{lscape}
\usepackage{setspace}
\usepackage{textcomp}
\usepackage{footmisc}
\usepackage [english]{babel}
\usepackage [autostyle, english = american]{csquotes}
\MakeOuterQuote{"}

\DeclareMathSizes{10}{9}{6}{6}

\begin{document}


\title{Neutron Phase Spin Echo}

\author{Florian M. Piegsa}
\email{florian.piegsa@phys.ethz.ch}
\affiliation{ETH Z\"urich, Institute for Particle Physics, CH-8093 Z\"urich, Switzerland}

\author{Patrick Hautle}
\affiliation{Paul Scherrer Institute, CH-5232 Villigen PSI, Switzerland}



\author{Christian Schanzer}
\affiliation{SwissNeutronics AG, CH-5313 Klingnau, Switzerland}

\date{\today}

\begin{abstract}
A novel neutron spin resonance technique is presented based on the well-know neutron spin echo method. 
In a first proof-of-principle measurement using a monochromatic neutron beam, it is demonstrated that relative velocity changes of down to a precision of $4 \times 10^{-7}$ can be resolved, corresponding to an energy resolution of better than 3~neV. Currently, the sensitivity is only limited by counting statistics and not by systematic effects. An improvement by another two orders of magnitude can be achieved with a dedicated setup, allowing for energy resolutions in the 10~peV regime. The new technique is ideally suited for investigations in the field of precision fundamental neutron physics, but will also be beneficial in scattering applications.
\end{abstract}




\pacs{61.05.F-, 03.75.Be, 78.70.Nx, 42.30.Rx}








\maketitle

\section{INTRODUCTION}
The neutron spin echo (NSE) technique was first proposed by Mezei in 1972 \cite{Mezei1972}. 
It describes a very sensitive method to determine small changes of the neutron velocity caused by inelastic scattering in a sample \cite{MezeiPappas2003}. 
It can be equally well used to investigate effects acting on the neutron spin itself, e.g.\ spin-dependent scattering, magnetic interactions etc. \cite{Zimmer2002,Frank2002,Kindervater2012}. Nowadays, NSE beam lines are part of the instrument suite of almost every major neutron scattering facility world-wide as they provide excellent energy resolutions and often deliver unique and complementary information compared to other scattering techniques \cite{Farago2015,Farago1999,Schleger1999,Holderer2008,Hino2013,Ohl2012,Pasini2015,Zanotti2011,Rosov1999}. Moreover, NSE and related neutron spin precession techniques, e.g.\ Ramsey's technique of separated oscillatory fields, are widely used in fundamental neutron physics, where one usually investigates tiny spin-dependent interactions or small energy transfers \cite{Ramsey1949,Ramsey1950,Ramsey1986, Dubbers2011, Piegsa2014, Pignol2015}. \\
Like its counterpart, the pulse nuclear magnetic resonance (NMR) method \cite{Hahn1953}, the NSE sequence consists of a series of pulses acting on the neutron spins. In between these pulses, i.e.\ spin flips, the spins precess freely in the plane perpendicular to a constant externally applied magnetic field $B_0$.
In a classic NMR Hahn echo scheme only a $\pi/2$- and a $\pi$-pulse, separated by a time $T$, are applied and the reoccurring echo of previously dephased spins induces a signal in a pick-up coil \cite{Slichter1996}. In NSE a second $\pi/2$ pulse needs to be applied after a second waiting time $T$ and the final neutron polarization along the quantization axis ($B_0$ direction) is detected using a neutron spin analyzer \cite{Williams1988}. 
An even more elaborate version of NSE is the so-called neutron resonance spin echo (NRSE) technique described by G\"ahler and Golub \cite{Gahler1987,Golub1987}. In this concept the spins are not flipped by static magnetic fields tilted with respect to the main field $B_0$, but instead one uses resonant high-frequency (HF) magnetic field pulses close to the Larmor precession frequency $\omega_0=-\gamma_{\text{n}} B_0$ in order to flip the neutron spins in and out of their plane of precession.\footnote{In such a scheme the precession field between the HF pulses is usually chosen to be close to zero, since this proves advantageous for neutron scattering within a large solid angle \cite{Haussler2011}.} Here, $\gamma_{\text{n}} = -2 \pi \times 29.165$~MHz/T is the gyromagnetic ratio of the neutron \cite{Greene1979}.   
Detailed theoretical descriptions of the spin flip process, NSE, and NRSE can be found in Ref.\ \cite{Golub1994,Ignatovich2003,Piegsa2015}.

\section{MEASUREMENT PRINCIPLE}
A conventional NSE setup is divided into two precession regions or arms, each with a length $L$. At the start of the first arm the spins of a polarized monochromatic neutron beam get flipped by $\pi/2$ into the precession plane perpendicular to the $B_0$ field. At the end of the first arm the neutron spins have accumulated a total  precession phase $\varphi = - \gamma_{\text{n}} B_0 L/v$, where $v$ is the neutron velocity. This velocity-dependence causes an overall dephasing (fanning out) of the spins. The dephasing is reversed by the $\pi$ flip applied half way through the NSE setup.
Hence, the spins are back in phase when they reach the second $\pi/2$ flipper at the end of the second arm. A so-called NSE signal is obtained by measuring the neutron count rate behind a spin analyzer either by scanning the magnetic field using auxiliary coils in one of the arms or by varying the length of one precession arm \cite{Dubbers1989,Habicht2004,Ioffe2004235}. The result is a sinusoidally oscillating count rate as a function of the scanning parameter. Due to the finite width of the employed neutron beam velocity distribution of typically 10-15\%, the oscillation pattern is damped with a spectrum dependent envelope function.
Small changes of the neutron velocity $\Delta v$, for instance due to inelastic scattering in a sample placed either right in front or behind the $\pi$ flipper, will result in a shift of the net phase of the neutron spins and eventually also of the phase of the NSE oscillation pattern by:
\begin{equation}
\Delta \varphi = \gamma_{\text{n}} B_0 L \cdot \frac{\Delta v}{v_0^2}
\label{eq:deltaphi1}
\end{equation}
with $v_0$ being the initial mean velocity of the neutron spectrum. For small relative velocity changes, i.e.\ $\Delta v/v_0 \ll 1$, the corresponding relative kinetic energy change is related via $\Delta E / E_0 \approx 2 \cdot \Delta v/v_0$, where $E_0 = \frac{1}{2} m v_0^2$ and $m$ is the neutron mass. Thus, measuring the phase shift $\Delta \varphi$ represents a direct method to determine the change of the kinetic energy:
\begin{equation}
\Delta E = \frac{m v_0^3}{\gamma_{\text{n}} B_0 L} \cdot \Delta \varphi
\label{eq:deltaevhoch3}
\end{equation}
From this equation it becomes obvious that there are three ways to improve the energy resolution of a NSE instrument. Firstly, by increasing the field integral $B_0 L$, which is limited by the technical difficulty to provide large homogeneous magnetic fields and the total length of the setup. Secondly, by decreasing the neutron velocity, however, this is accompanied by a strongly decreasing neutron beam intensity due to the shape of the Maxwell velocity distribution. Finally, the option followed in the present work, which relies on reducing the uncertainty $\sigma(\Delta \varphi)$ of the phase determination. 
\begin{figure}
	\centering
		\includegraphics[width=0.45\textwidth]{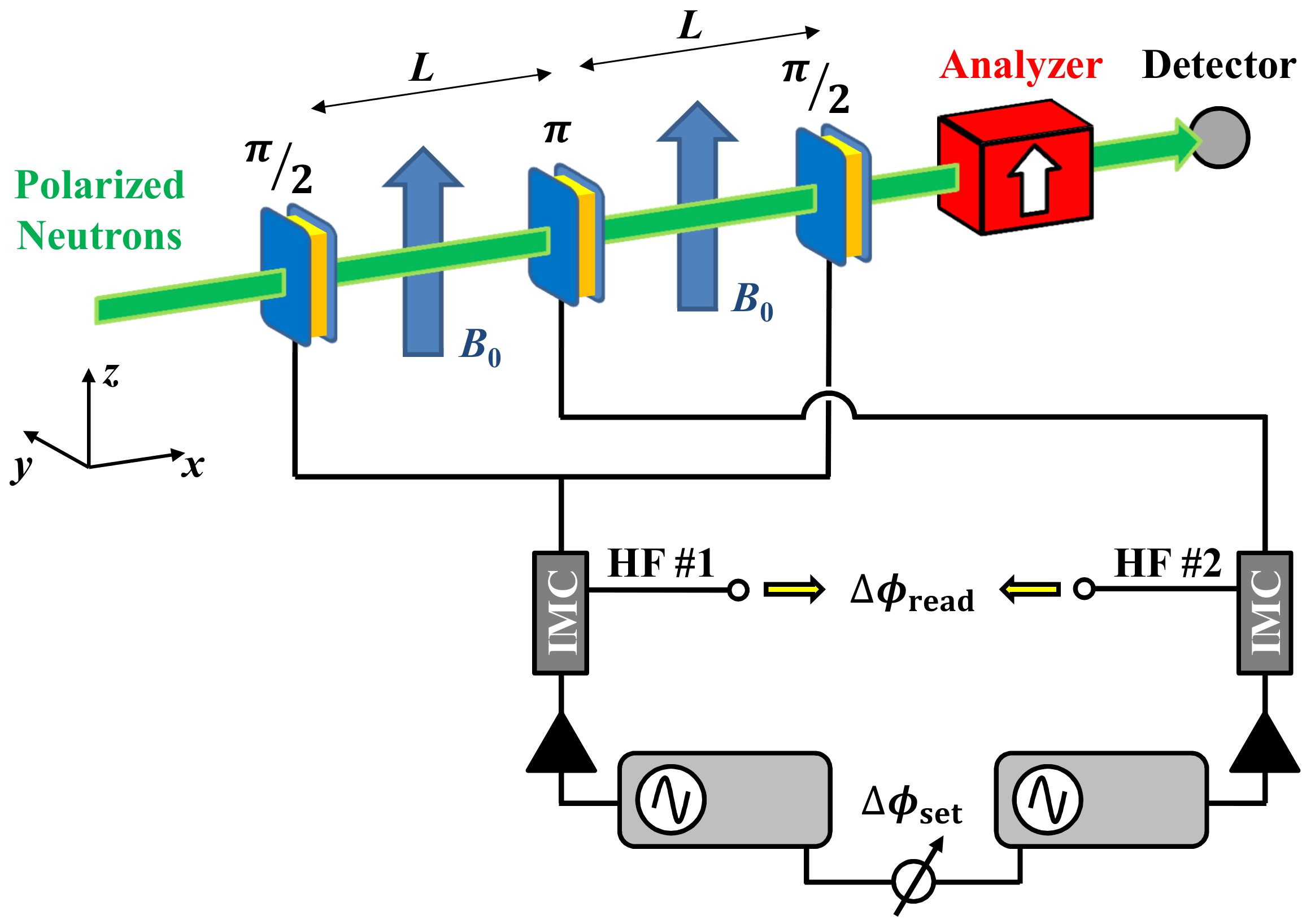}
		\caption{(color online). Scheme of the neutron phase spin echo setup. Compare text for details.  }
	\label{fig:Fig1_Setup}
\end{figure}
\begin{figure}
	\centering
		\includegraphics[width=0.40\textwidth]{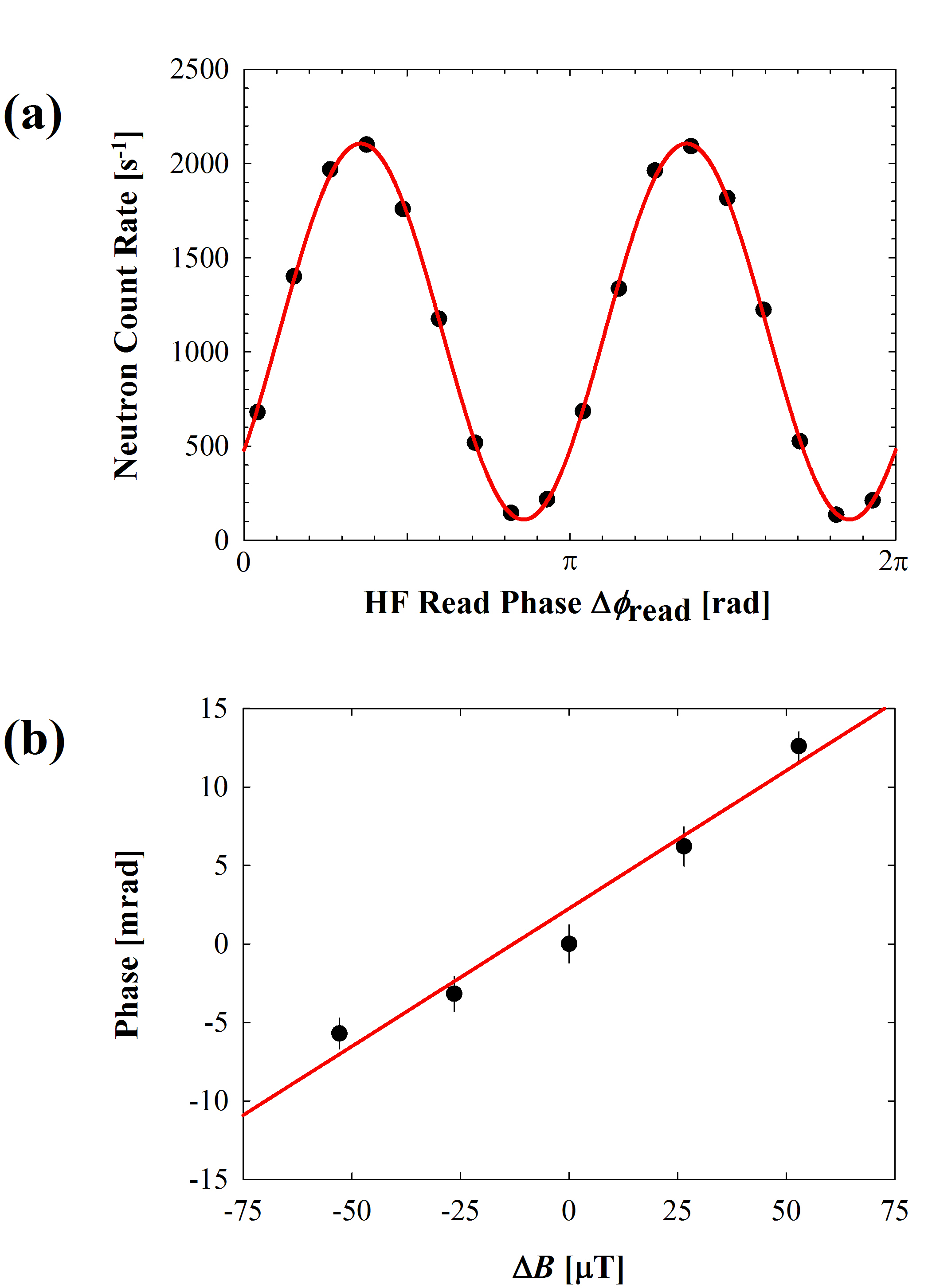}
	\caption{(color online). (a) Example of a phase scan as a function of $\Delta \phi_{\text{read}}$. Each data point was measured for 2~s, corresponding to a total scan time of about 40~s. The curve represents a sinusoidal fit to the data points. Assuming a fixed period of $\pi$ the phase retrieval uncertainty is approximately 7~mrad in such a single scan. The signal visibility reaches about 88\% in agreement with the analyzing powers of the spin polarizer/analyzer. (b) Measured susceptibility of the NPSE signal phase on detuning the external magnetic field $B_0 = 27.8$~mT by $\Delta B$ and corresponding linear fit (line).}
	\label{fig:Example}
\end{figure}
\begin{figure*}
	\centering
		\includegraphics[width=0.95\textwidth]{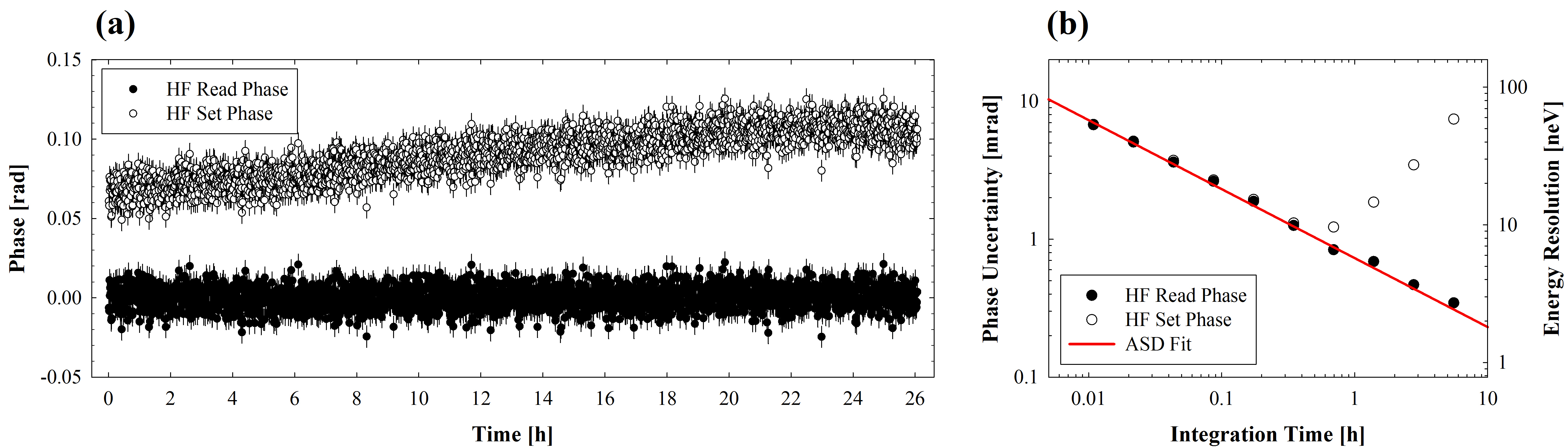}
	\caption{(color online). (a) Phase stability run of the NPSE signals over a period of approximately 26~h corresponding to 2400 single measurements. The phases are determined employing the values for the set $\Delta \phi_{\text{set}}$ and the read $\Delta \phi_{\text{read}}$ HF phase. The plot has been shifted such that the average phase using the $\Delta \phi_{\text{read}}$ values coincides with zero. (b) Corresponding ASD plot of the phase uncertainty $\sigma (\Delta \varphi )$ and the energy resolution as a function of the integration time $t$. Fitting the ASD data (line) with the function $f(t) = a / \sqrt{t}$ yields $a \approx 44$~mrad$/\sqrt{\text{Hz}}$ or 
	$a \approx 345$~neV/$\sqrt{\text{Hz}}$, respectively.}
	\label{fig:Fig3_StabilityASD}
\end{figure*}\\
The gist of the novel neutron phase spin echo (NPSE) technique is based on the standard NRSE method, however, aiming for high precision and stability of the magnetic field $B_0$ and the entire setup in order to reach optimum sensitivity. This is achieved in a most reproducible and accurate manner if the data is not taken as a function of a magnetic field variation or of a mechanical movement of one of the spin flippers, but instead by scanning the phase of the HF signals feeding the spin flip coils. This has the eminent advantage that the spin flippers can be continuously operated on perfect resonance, i.e.\ with their common oscillation frequency 
$\omega$ equal to the neutron Larmor frequency $\omega_0$. In consequence, the magnetic field can be monitored and actively stabilized very precisely. 
In principle, the signal phase of any of the three spin flippers can be scanned with respect to the other two. However, since the signal amplitudes of the $\pi$-flipper and the two $\pi/2$-flippers are different, it is also reasonable to scan their relative phases. A scheme of the NPSE setup is presented in Fig.\ \ref{fig:Fig1_Setup}.
Two identical phase-locked signal generators with a tunable phase difference $\Delta \phi_{\text{set}}$ deliver input signals for two HF amplifiers connected to impedance matching circuits (IMC) and the corresponding spin flip coils. The IMC are equipped with bypass outputs (HF \#1 and \#2) which are used to continuously monitor the actual relative signal phase $\Delta \phi_{\text{read}}$ in order to correct for possible phase drifts in the amplifiers and the IMC. 
The three coils are separated by a distance $L$ and placed in a homogeneous static magnetic field $B_0$ in $z$-direction. 
A neutron beam initially polarized along the $z$-axis passes through the three spin flip coils. In the coils the neutrons experience a longitudinal ($x$-direction) oscillating HF field causing the neutron spins to flip by $\pi/2$ and $\pi$ depending on the magnitude of the oscillating current.
Behind the last coil a neutron spin analyzer is placed in front of a detector.

\section{EXPERIMENTAL APPARATUS AND RESULTS}
\begin{figure}
	\centering
		\includegraphics[width=0.40\textwidth]{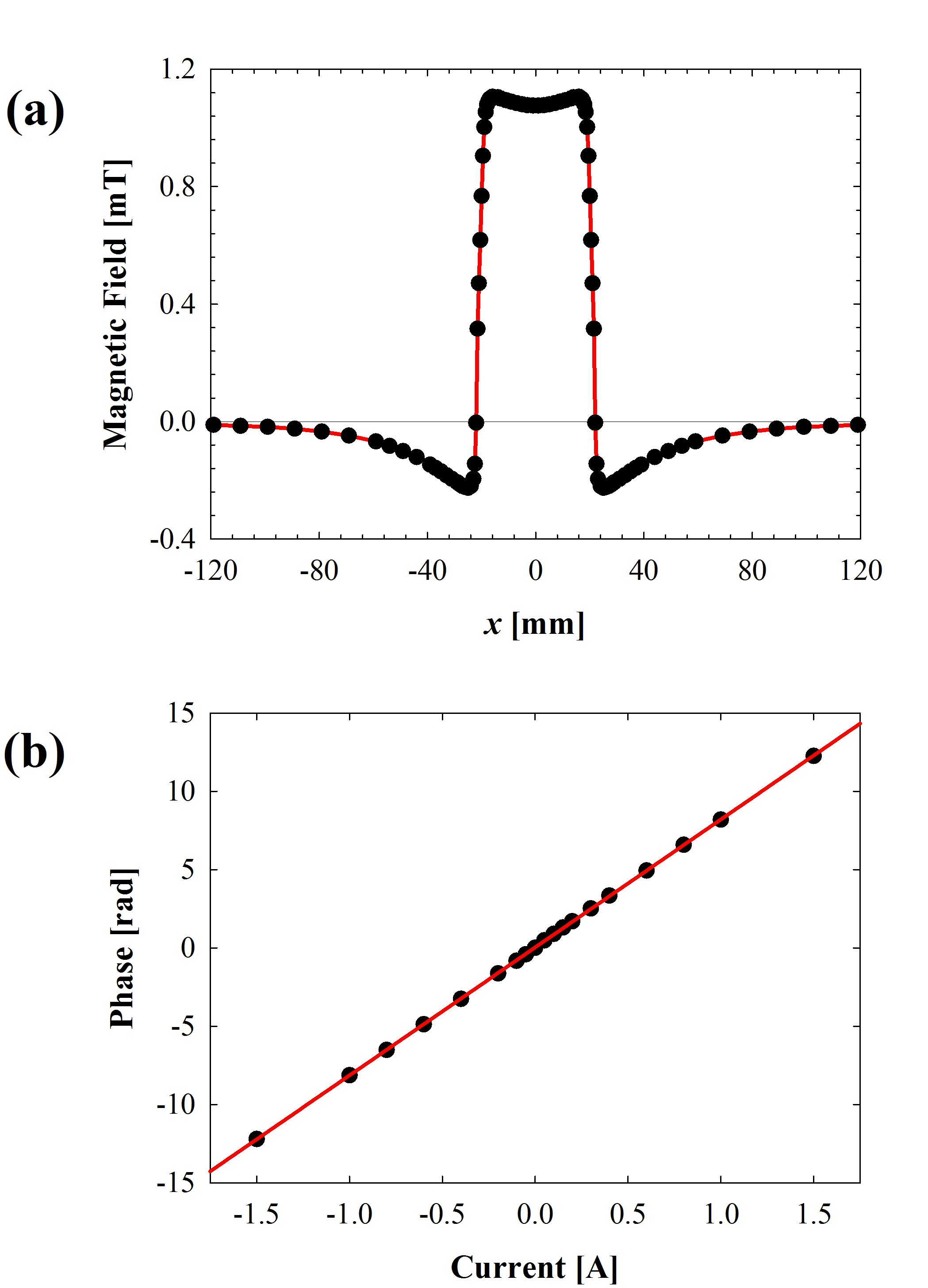}
	\caption{(color online). (a) Measured field of the test-coil in $z$-direction along the neutron flight path ($x$-axis) for a current of 1~A. The curve represents an interpolation through the data points. (b) Phase shift of the NPSE signal for different currents in the test-coil and linear fit (line) with a slope of $(8.17 \pm 0.01)$~rad/A. }
	\label{fig:Fig4_Coil}
\end{figure}

For a first proof-of-principle measurement, such a NPSE setup was installed at the reflectometer beam line Narziss at the spallation neutron source SINQ at the Paul Scherrer Institute (Switzerland). The monochromator of the reflectometer provides a neutron beam with a wavelength of $0.5$~nm, i.e.\ a velocity 
$v_0=791$~m/s, and a wavelength spread of 1.5\%. Two adjustable collimating slits in front of the homogeneous magnetic field region (separated by 725~mm) and two slits in front of the detector (separated by 995~mm) are all set to a width of 2~mm and a height of 40~mm. Two polarizing supermirrors, each with an analyzing power of better than 95\%, are situated between the two slit pairs and are used to polarize and analyze the neutron spins in the $z$-direction, respectively. The neutrons are counted using a $^3$He gas detector with an experimentally determined dead-time of 8~\textmu s.
The magnetic field $B_0$ is produced by the 60~cm-long instrument magnet which is adjusted to provide a vertical/transverse field of about 27.8~mT. The latter corresponds to a neutron Larmor frequency of 810~kHz, which is chosen as the driving frequency $\omega/(2 \pi)$ of the signal generators. 
The magnet itself has intrinsically a good stability, however, it is not actively stabilized or monitored with magnetic field sensors.
The three identical HF coils with a cross section of $30 \times 70$~mm$^2$ and of length 30~mm in $x$-direction are covered on both ends with 2~mm thick aluminum plates to effectively suppress the HF fringe fields \cite{Piegsa2015}. Each coil consists of 55 windings of a 0.5~mm thick copper wire. The distance between the spin flip coils is $L=130$~mm. This yields a total field integral of only $B_0 L \approx 3.6 \times 10^{-3}$~Tm, which is about two-to-three orders of magnitude smaller than at conventional NSE instruments. The IMC consist of a power resistor (4.7~$\Omega$) and a high-voltage capacitor ($0.25-0.5$~nF) forming together with the spin flip coils a serial RCL-circuit.
The bypass HF signals are digitized to determine $\Delta \phi_{\text{read}}$ by means of a scope card with a sampling rate of 10~MS/s and 8~bit resolution. 
An example of a NPSE signal is presented in Fig.\ \ref{fig:Example}a. 
The plot shows the expected period of $\pi$ and not of $2 \pi$, since the phase of the center spin flip coil is scanned. The maximum count rates are not located at $\Delta \phi_{\text{read}} = 0$, $\pi$, and $2 \pi$, as the magnetic field is not perfectly homogeneous along the neutron beam path. However, this is insignificant, 
since only relative phase shifts of the NPSE signal or changes in the signal visibility are investigated.\footnote{A variation of the NPSE signal visibility can be of interest in certain types of measurements, for instance in quasi-elastic scattering.} \\
The crucial property of any NSE system is the fact that it compensates global field drifts to a large extent. To investigate the susceptibility of the NPSE setup, the magnetic field was intentionally scanned close to $B_0$ by a small amount $\left|\Delta B\right| \leq  50$~\textmu T. The resulting phase shift measurement is shown in Fig.\ \ref{fig:Example}b. A linear fit to the data yields a shift of $\left(\frac{\Delta \varphi}{\Delta B}\right)= (0.175 \pm 0.012)$~mrad/\textmu T. This measured value is used to determine the dimensionless suppression factor $\xi$, defined by:
\begin{equation}
  \xi = \frac{\gamma_{\text{n}} L}{v_0} \cdot \left(\frac{\Delta \varphi}{\Delta B}\right)^{-1}
	\label{eq:suppressionfactor}
\end{equation}
This leads to $\xi = 172 \pm 12$, which is probably mainly limited by the mechanical accuracy of the setup (e.g.\ the positioning of the spin flip coils) and the homogeneity of the magnetic field.\\
The stability and hence the sensitivity of the apparatus is deduced by repeatedly taking NPSE signals and determining the phase of their oscillation patterns. The result of such a cycled measurement carried out over a period of about 26~hours is presented in Fig.\ \ref{fig:Fig3_StabilityASD}. 
The sensitivity is best described by a so-called Allan standard deviation (ASD) plot which is usually employed to characterize the sensitivity of systems like atomic clocks as a function of measuring/integration time \cite{Allan1966}.
The phases of the individual NPSE signals are analyzed with a sinusoidal fit (compare Fig.\ \ref{fig:Example}a), by either using the set HF phase $\Delta \phi_{\text{set}}$ or the monitored phase $\Delta \phi_{\text{read}}$ values. The phase obtained with the $\Delta \phi_{\text{set}}$ values still exhibits a drift of several mrad over the entire measurement time. This is probably due to drifts in the HF amplifiers and the IMC.  
The phase determined using the $\Delta \phi_{\text{read}}$ values, however, shows no systematic deviation from the expected statistically improving sensitivity.
This is also visible in the corresponding ASD plot, where a sensitivity of about 0.35~mrad is reached after 5.6~hours (512 single NPSE scans). Using Eq.\ (\ref{eq:deltaevhoch3}), this corresponds to a remarkable energy resolution of $\Delta E=2.7$~neV and with $E_0 = 3.3$~meV to a relative energy and velocity sensitivity of 
$\Delta E / E_0 =  8 \times 10^{-7}$ and  $\Delta v / v_0 = 4  \times 10^{-7}$, respectively.\footnote{One can define a corresponding time resolution 
$\Delta t= \sigma (\Delta \varphi )/\omega_0$. This yields $\Delta t \approx 70$~ps, with $\sigma (\Delta \varphi )=0.35$~mrad. } It is important to point out that the sensitivity is so far only limited by the neutron statistics and not by the stability of the magnetic field or by other components of the apparatus.\\
To demonstrate the capability of the NPSE setup, the phase shift induced by the magnetic interaction of the neutrons with an additional external field is measured.
The expected magnetic phase shift is given by:
\begin{equation}
  \varphi_{\text{magn}} = \frac{\gamma_{\text{n}}}{v_0} \cdot \int \Delta B_z(x) \text{ } dx
	\label{eq:magnphaseshift}
\end{equation}
where $\Delta B_z(x)$ describes the field change in $z$-direction, e.g.\ caused by an auxiliary coil. 
In this example measurement an approximately 40~mm thick ($x$-direction) box shaped test-coil was introduced in the first arm of the NPSE setup. The coil produces a magnetic field parallel to $B_0$ and has a length of 70~mm ($z$-direction), thus, producing a rather homogeneous field over the entire neutron beam cross section. The magnetic field as a function of the $x$-position was measured by means of a Hall probe (compare Fig.\ \ref{fig:Fig4_Coil}a). 
This information is used to determine the total field integral per coil current seen by the neutrons of $(35 \pm 1) \times 10^{-6}$~Tm/A, which yields an expected additional spin precession angle of $(8.1 \pm 0.2)$~rad/A using Eq.\ (\ref{eq:magnphaseshift}). This value is in good agreement with the result of $(8.17 \pm 0.01)$~rad/A obtained from performing NPSE measurements at several different test-coil currents presented in Fig.\ \ref{fig:Fig4_Coil}b.
 \\
In order to further improve the sensitivity of the presented NPSE apparatus the magnetic field $B_0$ and the length $L$ can be increased to 0.1~T and 0.5~m, respectively. This yields a total field integral of $50\times 10^{-3}$~Tm, which is about a factor 14 larger than in the current setup. By allowing for a broader velocity band and going to a stronger neutron source or beam line another statistical improvement factor of $\sqrt{50}$ is realistic. Hence, without even changing the mean neutron velocity $v_0$, an absolute energy sensitivity of 30~peV and a relative energy resolution of $\Delta E /E_0 \approx 10^{-8}$, i.e.\ $\Delta v /v_0 \approx 5 \times 10^{-9}$, seems possible. 
However, this requires a more stringent consideration of the magnetic field stability and homogeneity. From Eq.\ (\ref{eq:deltaevhoch3}) and (\ref{eq:suppressionfactor}), a limit on the relative stability of the magnetic field is deduced:
\begin{equation}
	\frac{\Delta B }{B_0} < \frac{\xi}{2} \cdot\frac{\Delta E}{E_0}
\end{equation}
With an assumed suppression factor $\xi$ of 200 and the aforementioned relative energy resolution, this requires a magnetic field stability of $10^{-6}$. 
In addition, to avoid a decrease in the NPSE signal visibility, the field homogeneity over the beam cross section must fulfill: 
\begin{equation}
  \delta B_{\text{homg}} \ll \frac{2\pi}{\gamma_{\text{n}}} \cdot \frac{v_0}{L}
\end{equation}
This yields an absolute homogeneity of $\delta B_{\text{homg}} \ll 55$~\textmu T or a relative homogeneity of better than $5\times 10^{-4}$.
Both requirements seem achievable by applying active stabilization with state-of-the-art Hall or NMR probes and by performing a careful magnet design. 
An additional option consists in introducing the so-called two beam method, where a second neutron beam is used as a reference to compensate for potential common field/phase drifts \cite{Zimmer2002,Piegsa2008b}. 

\section{CONCLUSION}
In conclusion, the presented NPSE technique represents an ideal tool and offers a wealth of new possibilities in precision fundamental neutron physics. However, due to its large gain in stability and reproducibility, compared to conventional NSE, the method will also proof beneficial in neutron scattering applications.
Examples of fundamental neutron physics experiments encompass the measurement of the neutron electric dipole moment with a pulsed beam \cite{Piegsa2013a}, the search for new exotic interactions mediated by axion-like particles \cite{Yan2013,Piegsa2012a}, the search for parity-violating neutron spin rotations \cite{Snow2011}, 
the measurement of the fine-structure constant \cite{Kruger1995},
observation of neutron acceleration due to gravity by extending the NPSE method to very cold or ultracold neutrons \cite{Golub1991}, and the measurement of incoherent neutron scattering lengths \cite{Abragam1975,Glattli1979,Zimmer2002,Piegsa2008b}. Moreover, the NPSE technique can also lead to a boost in sensitivity in polarized neutron radiography \cite{Kardjilov2008,Strobl2009,Piegsa2008a,Piegsa2009a,Piegsa2011a,Treimer2014,Tremsin2015}. Finally, a setup which combines NPSE with Ramsey's method of separated oscillatory fields would be ideal to distinguish between effects that only act on the neutron spin, e.g.\ magnetic interaction, and effects that change the neutron velocity, e.g.\ optical potential/refractive index.

\section{ACKNOWLEDGMENTS}
The authors are grateful for many fruitful discussions with Peter B\"oni, Klaus Kirch, Andreas Michels, and Ivan Titov.
This work has been performed at the Swiss Spallation Neutron Source SINQ at the Paul Scherrer Institute in Villigen, Switzerland.






\appendix
\bibliography{piegsa-bibfile}
\end{document}